\documentclass[11pt]{amsart}

\numberwithin{equation}{section}

\usepackage{color}
\usepackage{mathptmx}       
\usepackage{helvet}         
\usepackage{courier}        
\usepackage{graphicx, graphics, subfigure}        

\usepackage{amsmath}
\usepackage[sort]{cite} 

\usepackage{sidecap}
\sidecaptionvpos{figure}{c}

\usepackage[colorlinks=true, pdfstartview=FitV, linkcolor=blue,
            citecolor=blue, urlcolor=blue]{hyperref}  

\graphicspath{{../figures/}{./figures/}}

\newtheorem{rem}{Remark}[section]
\def\br{\begin{rem}}
\def\er{\end{rem}}

\def\d{\, \mathrm{d}}
\def\be{\begin{equation}}
\def\ee{\end{equation}}
\def\bes{\begin{equation*}}
\def\ees{\end{equation*}}
\def\bea{\begin{equation} \begin{aligned}}
\def\eea{\end{aligned} \end{equation}}
\def\beas{\begin{equation*} \begin{aligned}}
\def\eeas{\end{aligned} \end{equation*}}

\definecolor{rred}{rgb}{0.7,0,0.1}
\definecolor{greenrb}{rgb}{0.2,0.6,0.2}

\newcommand{\oa}{\overline{a}}

\newcommand{\bu}{\boldsymbol{u}}
\newcommand{\obu}{\overline{\boldsymbol u}}

\newcommand{\bx}{\boldsymbol{x}}
\newcommand{\bX}{\boldsymbol{X}}
\newcommand{\bphi}{\boldsymbol{\varphi}}

\title[Regularized Reduced Order Models for a Stochastic Burgers Equation]{Regularized Reduced Order Models \\ for a Stochastic Burgers Equation}

\author[Traian Iliescu]{Traian Iliescu}
\address[TI]{Department of Mathematics, Virginia Tech, 225 Stanger Street, Blacksburg, Virginia 24061, USA} 
\email{iliescu@vt.edu}
\author[Honghu Liu]{Honghu Liu}
\address[HL]{Department of Mathematics, Virginia Tech, 225 Stanger Street, Blacksburg, Virginia 24061, USA} 
\email{hhliu@vt.edu}
\author[Xuping Xie]{Xuping Xie}
\address[XX]{Department of Mathematics, Virginia Tech, 225 Stanger Street, Blacksburg, Virginia 24061, USA} 
\email{xupingxy@vt.edu}

\keywords{Reduced-order modelling, Leray regularized model, stabilization method, numerical instability, stochastic Burgers equation, differential filter}

\subjclass[2010]{34F05, 35R60, 37L55, 60H15}
\begin{document}

\begin{abstract}
 
 In this paper, we study the numerical stability of reduced order models for convection-dominated stochastic systems in a relatively simple setting: a stochastic Burgers equation with linear multiplicative noise.
Our preliminary results suggest that, in a convection-dominated regime, standard reduced order models yield inaccurate results in the form of spurious numerical oscillations.
To alleviate these oscillations, we use the Leray reduced order model, which increases the numerical stability of the standard model by smoothing (regularizing) the convective term with an explicit spatial filter.
The Leray reduced order model yields significantly better results than the standard reduced order model and is more robust with respect to changes in the strength of the noise.

\end{abstract}

\maketitle

\section{Introduction}
\label{sec:introduction}

Reduced order models (ROMs) are commonly used in applications that require repeated numerical simulations of large, complex systems~\cite{hesthaven2015certified,quarteroni2015reduced}.
ROMs have been successful in the numerical simulation of various fluid flows~\cite{HLB96,noack2011reduced}.
Numerical instability, usually in the form of unphysical numerical oscillations, is one of the main challenges for ROMs of fluid flows described by the Navier-Stokes equations (NSE).
There are several sources of numerical instability of ROMs for fluid flows~\cite{caiazzo2013numerical}, such as (i) the convection-dominated (high Reynolds number) regime, in which the convection nonlinear term plays a central role~\cite{AHLS88,HLB96,noack2011reduced}; and (ii) the inf-sup condition, which imposes a constraint on the ROM velocity and pressure spaces~\cite{ballarin2015supremizer,caiazzo2013numerical}.
To mitigate the spurious numerical oscillations created by these sources of numerical instability, various stabilized ROMs have been proposed (see, e.g.,~\cite{amsallem2012stabilization,AHLS88,balajewicz2013low,ballarin2015supremizer,barone2009stable,bergmann2009enablers,carlberg2013gnat,cordier2010calibration,giere2015supg,kalashnikova2010stability,pacciarini2014stabilized,quarteroni2011certified,wang2012proper,xiao2014non} for such examples).
A promising recent development in this class of methods is regularized ROMs~\cite{sabetghadam2012alpha,wells2016regularized}, which use explicit spatial filtering to increase the numerical stability of the ROM approximation.


Recently, the development of ROMs for systems involving random components has also received increased attention. For instance, ROMs for partial differential equations (PDEs) subject to random inputs acting on the boundary as well as PDEs with random coefficients have been considered in various contexts \cite{Chen2015,Chen2016sparse,Galbally2010,Lassila2013reduced,Boyaval2010reduced,Boyaval2009reduced,Chen2013weighted,Haasdonk2013reduced,Torlo16}. However, ROMs for evolutionary PDEs driven by stochastic processes such as Brownian motions seem to be much less investigated. To our knowledge, only a few works are available \cite{burkardt2007reduced}; see also \cite{CLW15_vol2}, where a new stochastic parameterization framework is presented to address a related question of parameterizing the unresolved high-frequency modes in terms of the resolved low-frequency modes.



In this paper, we consider ROMs within the context of nonlinear stochastic PDEs (SPDEs) that are of relevance to fluid dynamics. The main purpose is to investigate within a simple relevant setting---a stochastic Burgers equation (SBE) driven by linear multiplicative noise---the stabilization of the standard Galerkin ROM (G-ROM) in a convection-dominated regime. It is numerically illustrated that spurious oscillations developed in a G-ROM persist as the noise is turned on, and the oscillations worsen as the noise amplitude increases. A Leray regularized ROM (referred to as L-ROM hereafter) is then tested. The L-ROM provides more accurate modeling of the SBE dynamics by greatly reducing the artificial oscillations of the G-ROM, especially when the dimension of the reduced models are low; cf.~Figs.~\ref{fig:questions-1-2-1}--\ref{fig:LROM_projection}. It is further illustrated that the L-ROM is much more robust than the G-ROM with respect to the noise amplitude as revealed by the statistics of the corresponding modeling errors, which have significantly lower mean and variance; cf.~Fig.~\ref{fig:question-4-1}.

The rest of the paper is organized as follows. In Section~\ref{sec:rom}, we outline the SBE to be used in our numerical exploration and derive the corresponding G-ROM and the L-ROM based on the proper orthogonal decomposition. The performance of the two ROMs is then tested and compared in Section~\ref{sec:computational-investigation} by placing the SBE in a convection-dominated regime. Finally, some concluding remarks and potential future research directions are given in Section~\ref{sec:conclusions}.


\section{Reduced Order Models for a Stochastic Burgers Equation}
	\label{sec:rom}

The viscous Burgers equation and its stochastic versions have been used previously to test new techniques in reduced order modeling and related contexts; see among many others \cite{CLW15_vol2,CLW16_additive,KV99,KV01,CL15,nguyen2009reduced}. In this paper, we will focus on a stochastic Burgers equation (SBE) driven by linear multiplicative noise, which is presented briefly in Section~\ref{sec:sbe}. To fix ideas, the ROMs explored in this paper will be derived based on the proper orthogonal decomposition (POD). In Section~\ref{sec:pod}, we outline the main steps in the derivation of the POD basis. The standard Galerkin ROM for the SBE is then derived in Section~\ref{sec:g-rom}. In Section~\ref{sec:l-rom}, we develop the Leray ROM, which is a regularized ROM that aims at increasing the numerical stability of the standard ROM for the SBE.


\subsection{Stochastic Burgers Equation (SBE)}  \label{sec:sbe}

In this paper, we focus on the following stochastic Burgers equation (SBE) driven by linear multiplicative noise:
\bea \label{eqn:sbe}
& \mathrm{d} u = \big( \nu u_{xx}  - u  u_x\big) \mathrm{d} t + \sigma u \circ \mathrm{d}W_t, \\
& u(0,t) = u(1,t) = 0, \;\;\; t\geq 0,\\
& u(x, 0) = u_0(x), \qquad x\in (0,1),
\eea
where $\nu$ is a positive diffusion coefficient, $W_t$ is a two-sided one-dimensional Wiener process, $\sigma$ is a positive constant which measures the ``amplitude" of the noise, and $u_0$ is some appropriate initial datum to be specified below.  To fix  ideas, the multiplicative noise term $\sigma u \circ \mathrm{d}W_t$ is understood in the sense of Stratonovich \cite{Oksendal03}.

SPDEs driven by linear multiplicative noise such as the SBE~\eqref{eqn:sbe}  arise in various contexts, including turbulence theory or non-equilibrium phase transitions \cite{Birnir13,Cross_al93,Mun04}, the modeling of randomly fluctuating environment  \cite{BM77} in spatially-extended harvesting models \cite{HSZ02,CR06,RC07,RC10,MMQ11,MMQ08}, or simply the modeling  of parameter disturbances \cite{Blomker07}.

\subsection{Proper Orthogonal Decomposition (POD)}
	\label{sec:pod}

We present in this section a very brief account of the proper orthogonal decomposition (POD). The reader is referred to, e.g.,~\cite{HLB96,noack2011reduced,Sir87abc} for more details.
The POD starts with the snapshots, which, in this paper, are numerical approximations of the SBE~\eqref{eqn:sbe} at different time instances.  
The POD seeks a low-dimensional basis that approximates the snapshots optimally with respect to a certain norm. 
In this paper, we employ the commonly used $L^2$-norm (see, e.g., \cite{KV01} for alternatives). 
The solution of the minimization problem is equivalent to the solution of an eigenvalue problem~\cite{caiazzo2013numerical}.
The POD subspace of a given dimension $r$ is spanned by the first $r$ POD basis functions, which are the normalized functions $\{ \bphi_{j} \}_{j=1}^{r}$ that correspond to the first $r$ largest eigenvalues of the aforementioned eigenvalue problem: 
\be
\bX^r := \text{span} \{ \bphi_1, \ldots, \bphi_r \}.
\ee
Note that the POD functions are orthogonal to each other with respect to the $L^2$-inner product $\langle \cdot , \cdot \rangle$ on the underlying phase space:
\be \label{eqn:POD_orth}
\langle \bphi_i, \bphi_j \rangle = \delta_{ij},
\ee
where $\delta_{ij}$ denotes the Kronecker-delta. Note also that in \eqref{eqn:POD_orth} and the remainder of the paper,  the POD basis functions are considered as continuous functions on the spatial domain, since they are linear combinations of finite element basis functions.

\subsection{Galerkin ROM (G-ROM) for SBE}
    \label{sec:g-rom}

The derivation of the POD-based Galerkin ROM (G-ROM) follows the standard Galerkin approximation procedure with the underlying basis taken to be the POD basis. For the sake of clarity, we sketch this derivation for the SBE \eqref{eqn:sbe} below.  
 Given a positive integer $r$, the $r$-dimensional POD Galerkin approximation $u_r$ of the SBE solution $u$ takes the following form:
\begin{equation}    
	{\bu}_r(\bx,t; \omega) 
 	:= \sum_{j=1}^r a_j(t; \omega) \bphi_j(\bx),
	\label{eqn:g-rom-1}
\end{equation} 
where the time-varying random coefficients $\{a_{j}(t, \omega)\}_{j=1}^{r}$ are determined by solving:
\be  \label{eqn:g-rom-1b}
\big \langle \mathrm{d} u_r, \bphi_j \big \rangle  = \big \langle \big( \nu (u_r)_{xx}  - u_r  (u_r)_x\big), \bphi_j \big \rangle \mathrm{d} t + \sigma \langle u_r, \bphi_j \rangle  \circ \mathrm{d}W_t, \qquad j = 1, \cdots, r.
\ee 
The above system can be recast into the following more explicit form by using the expansion of $u_r$ given in \eqref{eqn:g-rom-1} and the orthogonality property satisfied by the POD basis functions given in \eqref{eqn:POD_orth}:
\bea \label{eqn:g-rom-2}
\mathrm{d} a_j = \Big [ - \nu  \sum_{k = 1}^r  a_k \big \langle \big( (\bphi_k)_{x}, (\bphi_j)_{x} \big \rangle   + \sum_{k,l = 1}^r a_k a_l   \big \langle \bphi_k (\bphi_l)_x, \bphi_j \big \rangle \Big] \d t + \sigma a_j  \circ \d W_t,
\eea
where $j = 1, \cdots, r$. This system of stochastic differential equations (SDEs) is the $r$-dimensional Galerkin ROM for the SBE \eqref{eqn:sbe}.


\subsection{Leray ROM (L-ROM) for SBE}
	\label{sec:l-rom}

To investigate fixes for G-ROM's potential numerical instability in the convection-dominated regime of the SBE~\eqref{eqn:sbe}, we draw inspiration from the deterministic case and consider {\it regularized ROMs (Reg-ROMs)}. 
These Reg-ROMs belong to the wide class of stabilized ROMs (see, e.g.,~\cite{amsallem2012stabilization,AHLS88,balajewicz2013low,ballarin2015supremizer,barone2009stable,bergmann2009enablers,carlberg2013gnat,cordier2010calibration,giere2015supg,HLB96,kalashnikova2010stability,noack2011reduced,pacciarini2014stabilized,quarteroni2011certified,wang2012proper,xiao2014non} for such examples).
What distinguishes the Reg-ROMs from the other stabilized ROMs is that they increase the numerical stability of the model by using {\it explicit spatial filtering}, which is a relatively new concept in the ROM field~\cite{wang2012proper,sabetghadam2012alpha,wells2016regularized}.
In this study, we will use the simplest such Reg-ROM, the {\it Leray ROM (L-ROM)}~\cite{sabetghadam2012alpha,wells2016regularized}, which is based on a specific way of filtering the convective term in the SBE~\eqref{eqn:sbe} as explained below.

The Leray model was first used by Leray~\cite{leray1934sur} as a theoretical tool to prove local existence and uniqueness of weak solutions of the NSE.
The Leray model has been used as a numerical tool in the simulation of convection-dominated deterministic flows with standard (e.g., finite element) numerical methods~\cite{cheskidov2005leray,geurts2003regularization,layton2012approximate}. 
It has also been used to derive Reg-ROMs for deterministic systems in~\cite{sabetghadam2012alpha,wells2016regularized}.

The extension of the L-ROM proposed in~\cite{sabetghadam2012alpha,wells2016regularized} to the stochastic problem \eqref{eqn:sbe} at hand is straightforward. There is only one crucial difference in its derivation compared to the derivation of the G-ROM as outlined in Section~\ref{sec:g-rom}, which consists of replacing the nonlinear term $u_r (u_r)_x$ in \eqref{eqn:g-rom-1b} by a regularized term $\obu_r (u_r)_x$ here. This regularized version, $\obu_r$, of $u_r$ is obtained based on the usage of the following {\it POD differential filter (DF)}\footnote{Differential filters have been used in the simulation of convection-dominated flows with standard numerical methods~\cite{germano1986differential-b,germano1986differential}. 
In reduced order modeling, the DF was first used in~\cite{sabetghadam2012alpha} and extended in~\cite{wells2016regularized}.} : Let $\delta$ be the radius of the DF. For a given $\bu_r \in \bX^r$, find $\obu_r \in \bX^r$ such that
\begin{eqnarray}
	\big \langle
		\left(
			I - \delta^2 \Delta
		\right) \obu_r , \bphi_j
	\big \rangle
	= \langle \bu_r, \bphi_j\rangle,
	\quad \forall \, j=1, \ldots r \, .
	\label{eqn:df}
\end{eqnarray}

Namely, the $r$-dimensional L-ROM approximation $u_r$ of the SBE solution $u$ takes the following form:
\begin{equation}    
	{\bu}_r(\bx,t; \omega) 
	:=  \sum_{j=1}^r a_j(t; \omega) \bphi_j(\bx),
	\label{eqn:l-rom-0}
\end{equation} 
where the time-varying random coefficients $\{a_{j}(t, \omega)\}_{j=1}^{r}$ are determined by solving:
 \be  \label{eqn:l-rom-1}
\big \langle \mathrm{d} u_r, \bphi_j \big \rangle  = \big \langle \big( \nu (u_r)_{xx}  - \overline{u}_r  (u_r)_x\big), \bphi_j \big \rangle \mathrm{d} t + \sigma \langle u_r, \bphi_j \rangle  \circ \mathrm{d}W_t, \qquad j = 1, \cdots, r.
\ee 
Since at each time instance $t$, the sought regularization $\obu^r(t,\cdot; \omega)$ lives in $\bX^r$, it admits the following expansion:
\begin{equation}    
	{\obu}_r (t,x;\omega)
	\equiv \sum_{k=1}^r \oa_k(t;\omega) \bphi_k(\bx)  \, ,
	\label{eqn:l-rom-2}
\end{equation} 
where $\oa_j$  can be determined by using the expansion \eqref{eqn:l-rom-2} in \eqref{eqn:df}, which leads to
\begin{equation}    
	\sum_{k=1}^r \oa_k(t;\omega) \bphi_k
	= \overline{ \sum_{k=1}^r a_k (t;\omega) \bphi_k }  = \sum_{k=1}^r a_k (t;\omega) \overline{\bphi}_k \, ,
\end{equation} 
and the filtered POD mode $ \overline{\bphi}_k$, $1\le k \le r$, is determined via
\be
\big \langle  \left(I - \delta^2 \Delta \right) \overline{\bphi}_k , \bphi_j
	\big \rangle
	= \langle \bphi_k, \bphi_j\rangle,
	\quad \forall \, j=1, \ldots r \, .
	\label{eqn:df_phi}
\ee

Consequently, in contrast to the G-ROM given in \eqref{eqn:g-rom-2}, the $r$-dimensional L-ROM for SBE~\eqref{eqn:sbe} is given by:
\bea \label{eqn:l-rom-3}
\mathrm{d} a_j = \Big [ - \nu  \sum_{k = 1}^r  a_k \big \langle \big( (\bphi_k)_{x}, (\bphi_j)_{x} \big \rangle   + \sum_{k,l = 1}^r a_k a_l   \big \langle \overline{\bphi}_k (\bphi_l)_x, \bphi_j \big \rangle \Big] \d t + \sigma a_j  \circ \d W_t,
\eea
where $j = 1, \cdots, r$.


\section{Computational Investigation}
	\label{sec:computational-investigation}

In this section, we present a computational investigation on potential numerical instability of the standard G-ROM \eqref{eqn:g-rom-2} for the SBE~\eqref{eqn:sbe} and on a possible alleviation of such instability achieved by the L-ROM \eqref{eqn:l-rom-3}. 

It has been observed in a previous study \cite{wells2016regularized} that, for the deterministic Burgers equation placed in a convection-dominated regime, the G-ROM yields excessive spurious oscillations, especially when the dimension of the G-ROM is low. Similar to \cite{wells2016regularized}, we set up the numerical experiments for the SBE~\eqref{eqn:sbe} in a regime with a small diffusion coefficient ($\nu = 10^{-3}$) and a steep internal layer; see Section~\ref{sec:numerical_setup}. In Section~\ref{sec:questions-1-2}, the emergence of such oscillations is confirmed in the current stochastic setting as well. The improvement achieved by the L-ROM in the form of significant reduction of the spurious oscillations is then presented in Section~\ref{sec:question-3}. Finally, some preliminary statistical tests are presented in Section~\ref{sec:question-4}, which also shows the robustness of the L-ROM with respect to the strength of the noise.


\subsection{Setup of the Numerical Experiments} \label{sec:numerical_setup}

In this section, we present a short description of the setup of the numerical experiments. 

\medskip
\noindent {\bf Numerical Discretization of the SBE.} The SBE~(\ref{eqn:sbe}) is solved by a semi-implicit Euler scheme as given in \cite[Section~6.1]{CLW15_vol2}.
For the reader's convenience, we briefly describe the numerical discretization below, and refer to \cite[Section~6.1]{CLW15_vol2} for more details.
We also refer the reader to \cite{AG06, BJ13,burkardt2007reduced,Hou_al06,JK11,LR04} for other numerical approximation schemes of nonlinear SPDEs.
 
At each time step the nonlinearity $uu_x = (u^2)_x/2$ and the noise term $\sigma u \circ \d W_t$ are treated explicitly, and the other terms are treated implicitly. 
The Laplacian operator is discretized using the standard second-order central difference approximation. The resulting semi-implicit scheme reads as follows:
\begin{eqnarray}  
u_{j}^{n+1} - u_{j}^{n} = 
 \Big( \nu \Delta_d u_{j}^{n+1} + \frac{\sigma^2}{2}u_{j}^{n} - \frac{1}{2} \nabla_d\big((u_{j}^{n})^2 \bigr)\Big)\Delta t  + \sigma \zeta_n   u_{j}^{n} \sqrt{\Delta t} \, ,
	\label{eqn:sbe-discrete}
\end{eqnarray} 
where $u_{j}^{n}$ is the discrete approximation of $u(j\Delta x, n \Delta t)$, $\Delta x$ the mesh size of the spatial discretization, and $\Delta t$ the time step.  
The discretized Laplacian $\Delta_d$  and the discretized spatial derivative $\nabla_d$   in~\eqref{eqn:sbe-discrete} are given by
\begin{equation*}  
\Delta_d u_{j}^{n}= \frac{u_{j-1}^{n} - 2u_{j}^{n} + u_{j+1}^{n}}{(\Delta x)^2}; \quad \nabla_d \big( (u_j^{n})^2\big) = \frac{ (u_{j+1}^{n})^2 - (u_j^{n})^2 }{\Delta x}, \quad j \in \{1, \cdots, N_x - 2\} \, .
\end{equation*}  
The boundary conditions in~\eqref{eqn:sbe-discrete} are $u_0^n=u^n_{N_x-1}=0$, where $N_x$ is the total number of grid points used for the discretization of the spatial domain $[0, 1]$.
The $\zeta_n$ in~\eqref{eqn:sbe-discrete} are random variables drawn independently from a normal distribution $\mathcal{N}(0,1)$. 
Note that the additional drift term $\sigma^2 u_{j}^{n}/2$ in the RHS of \eqref{eqn:sbe-discrete} is due to the conversion of the Stratonovich noise term $\sigma u\circ \mathrm{d}W_t$ into its  It\^o form. 
Throughout the paper, the simulations of the SBE~\eqref{eqn:sbe} are performed for $\Delta t=10^{-4}$ and $N_x = 1025$ so that $\Delta x \approx 9.8\times 10^{-4}$. 
The diffusion coefficient $\nu$ is set to be $0.001$. The values of the parameter $\sigma$ will be specified below.

\medskip
\noindent  {\bf Choice of the Initial Data.}  The initial condition is chosen to be a mollified and slightly shifted version of the step function used in~\cite{KV99}, which is given by 
\begin{equation} 
	\bu_0(x) = \int_{-\infty}^{\infty} \xi(y) \phi_{\epsilon}(x-y) \d y,  \qquad x \in [0,1].
	\label{eqn:initial-condition}
\end{equation}
Here, $\xi$ is the step function defined by $\xi(x) =1$ if $x\in (0.05, 0.55)$ and $\xi(x) = 0$ otherwise. The mollifier $\phi_{\epsilon}$ is given by $\phi_{\epsilon}(x) = \frac{1}{\epsilon}\phi(\frac{x}{\epsilon})$ with 
\bes
\phi(x) = \begin{cases}
C \exp\big(-\frac{1}{(1-x^2)}\big) & \text{ if $|x| < 1$}, \\
 0 & \text{otherwise}, 
 \end{cases}
\ees
and the normalization constant $C$ is chosen such that $\int_{-1}^1 \phi(x) \d x = 0$. Throughout our numerical experiments, the parameter $\epsilon$ in the mollifier $\phi_{\epsilon}$ is set to be $\epsilon = 0.01$.

The modification adopted here is mainly intended to enforce the compatibility of the initial and boundary condition at the left boundary point ($x=0$) and to avoid any potential regularity issues that may arise in our numerical discretization of the SBE in \eqref{eqn:sbe-discrete} due to the discontinuity in the step function. 

As will be seen below, by choosing such a step-function like initial profile and by setting the diffusion constant $\nu$ sufficiently small, the SBE exhibits interesting transient dynamics that will turn out to be a good laboratory to study the potential instability of the G-ROM; cf.~Fig.~\ref{fig:numerical-discretization-1}.

\medskip 
\noindent{\bf Numerical Integration of the ROMs.} The discretization of the G-ROM \eqref{eqn:g-rom-2} and the L-ROM \eqref{eqn:l-rom-3}  are carried out by using a standard Euler-Maruyama scheme (see, e.g., \cite[p.~305]{KP92}). For instance, the corresponding G-ROM discretization is given by: 
\bea  \label{eqn:g-rom-3}
a_{j}^{n+1} - a_{j}^{n} & =  \Big [ - \nu  \sum_{k = 1}^r  a^n_k \big \langle \big( (\bphi_k)_{x}, (\bphi_j)_{x} \big \rangle + \frac{\sigma^2}{2}a_{j}^{n} \\
&  \qquad    + \sum_{k,l = 1}^r a^n_k a^n_l   \big \langle \bphi_k (\bphi_l)_x, \bphi_j \big \rangle \Big] \Delta t + \sigma \zeta_n   a_{j}^{n} \sqrt{\Delta t} \, , \quad j = 1, \cdots, r,
\eea
where, as in \eqref{eqn:sbe-discrete}, $\zeta_n$ are random variables drawn independently from a normal distribution $\mathcal{N}(0,1)$, and $n = 1,\cdots, N$, with $N$ being the total number of time steps.

\subsection{G-ROM Results: Spurious Oscillations}
	\label{sec:questions-1-2}

In this section, we assess the performance of the G-ROM in its ability to reproduce the SBE's spatio-temporal field for a fixed noise amplitude $\sigma = 0.3$ and an arbitrarily fixed realization of the noise. The statistical relevance of the results presented in this section is confirmed in Section~\ref{sec:question-4}.  

\begin{figure}[h]
\begin{center}
	\includegraphics[height=0.4\textwidth,width=0.85\textwidth]{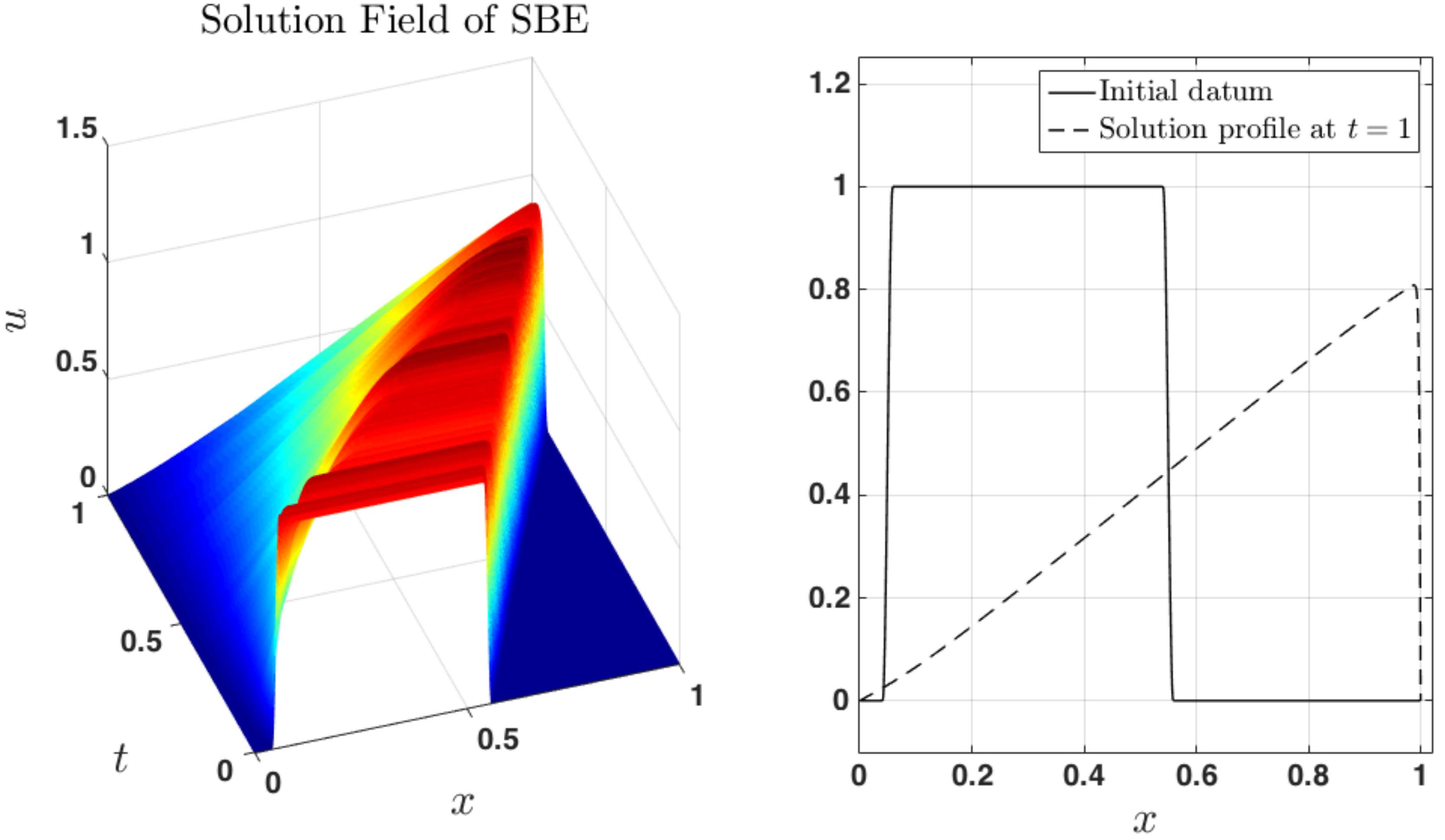}
	\caption{
		The numerically simulated spatio-temporal field of the SBE~\eqref{eqn:sbe} with $\sigma=0.3$ forced by an arbitrary realization of the noise (left panel), and the initial profile given by \eqref{eqn:initial-condition} with $\epsilon = 0.01$ (right panel, solid line) as well as the solution profile at time $t=1$ (right panel, dashed line).
	}
	\label{fig:numerical-discretization-1} 
	\end{center}
\end{figure}

For this purpose, we first simulate the SBE~\eqref{eqn:sbe} over the time interval $[0,1]$ following the numerical setup presented in Section~\ref{sec:numerical_setup} and construct the POD basis functions used in the derivation of the G-ROM~\eqref{eqn:g-rom-2}. 
In Fig.~\ref{fig:numerical-discretization-1}, the numerically simulated spatio-temporal field of the SBE~\eqref{eqn:sbe} as well as the initial profile and the final time solution profile are plotted.

\begin{figure} 
\begin{center}
	\includegraphics[height=0.35\textwidth,width=1\textwidth]{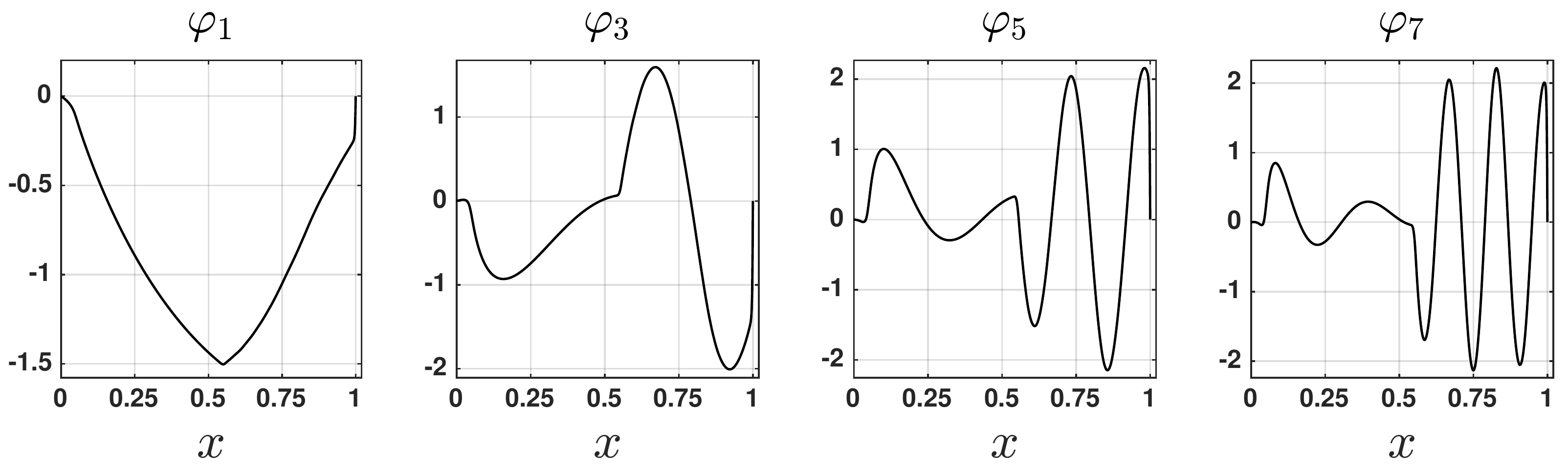}
	\caption{A few POD basis functions constructed based on the spatio-temporal field plotted in Fig.~\ref{fig:numerical-discretization-1}.  }
	\label{fig:numerical-discretization-2} 
	\end{center}
\end{figure}

To construct the POD basis functions used in the derivation of the G-ROM~\eqref{eqn:g-rom-2}, we collected $101$ equally spaced snapshots (without subtracting the centering trajectory) from the simulated SBE spatio-temporal field, and we used the method of snapshots~\cite{Sir87abc}. For illustration purposes, we plot four POD basis functions in Fig.~\ref{fig:numerical-discretization-2}.

\begin{figure}
\begin{center}
	\includegraphics[height=0.4\textwidth,width=1\textwidth]{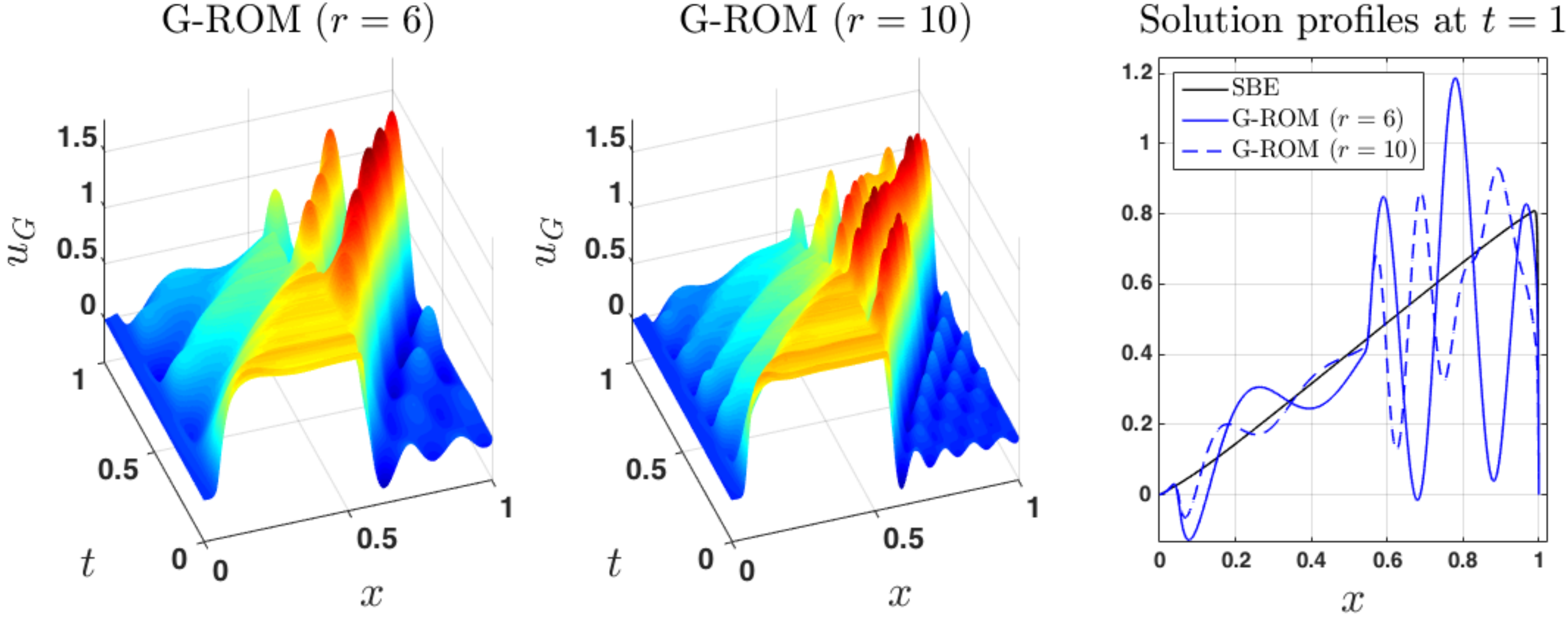}	
	\caption{
	Spatio-temporal field, $u_G:=\sum_{j=1}^r a_j \varphi_j$, reconstructed from the numerical simulation of the G-ROM~\eqref{eqn:g-rom-2} with dimension $r=6$ (left panel) and $r=10$ (middle panel), respectively. The noise path is the same as that used to generate the SBE's spatio-temporal field plotted in Fig.~\ref{fig:numerical-discretization-1}; $\sigma=0.3$.  The corresponding solution profiles at time $t=1$ are shown in the right panel. 
	}
	\label{fig:questions-1-2-1} 
	\end{center}
\end{figure}

\begin{figure} 
\begin{center}
	\includegraphics[height=0.5\textwidth,width=1\textwidth]{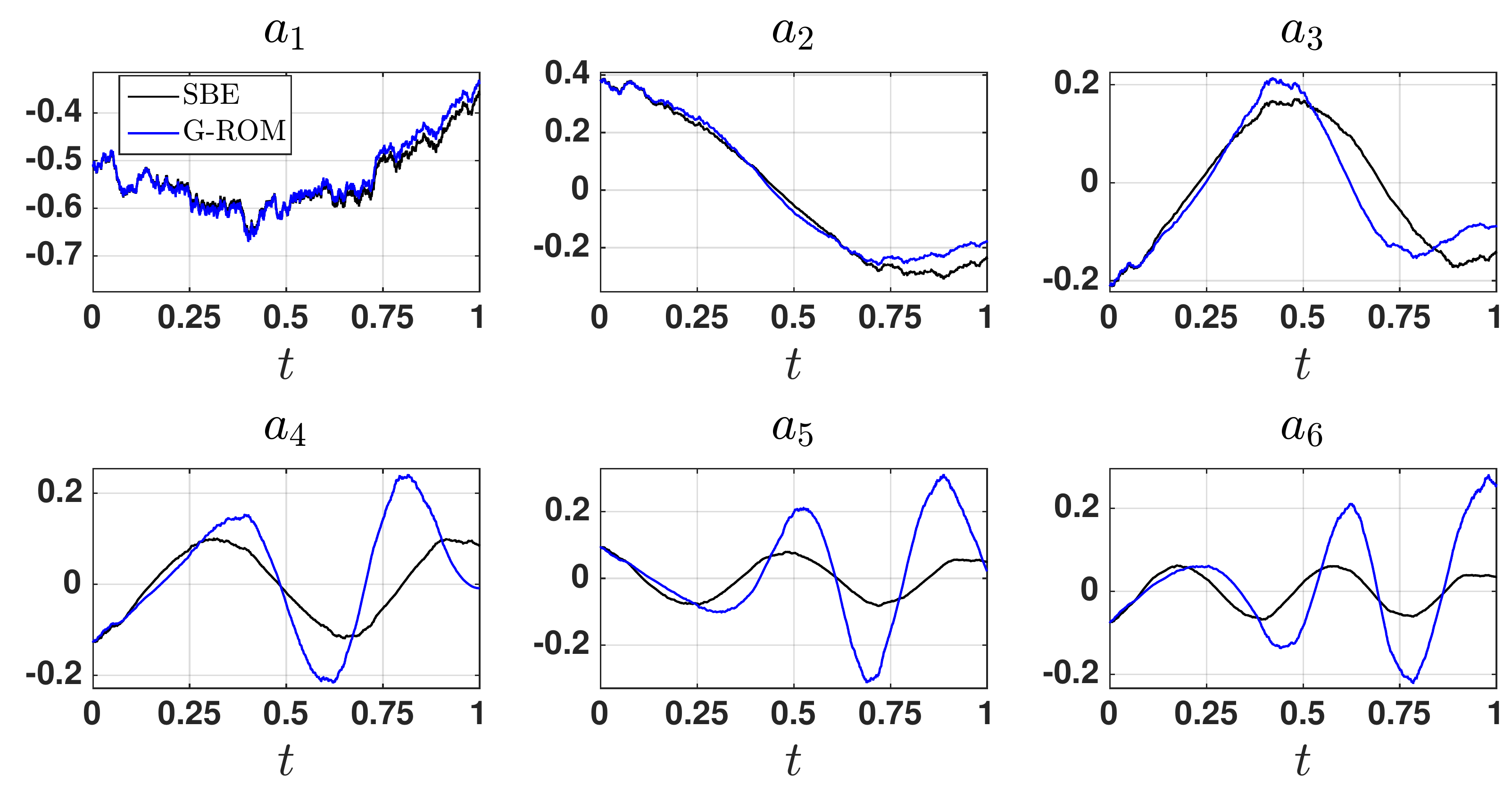}	
	\caption{
	The time series $a_j$, $1\le j \le r$, as modeled by the G-ROM \eqref{eqn:g-rom-2} with dimension $r=6$ (blue curves). Also plotted are the SBE solution projections onto the first $r$ POD modes (black curves).  
	}
	\label{fig:GROM_projection} 
	\end{center}
\end{figure}

The tests for the G-ROM are carried out with dimension $r=6$ as well as $r=10$; the results are plotted in Fig.~\ref{fig:questions-1-2-1}. In both cases, the percentage of the total kinetic energy contained in the first $r$ modes is already high: $98.5\%$ for $r=6$ and  $99.3\%$ for $r=10$. Despite such a high percentage of energy captured by the first $r$ POD modes, the corresponding G-ROM exhibits very strong spurious oscillations, as can be observed from both the reconstructed spatio-temporal fields and the final-time solution profiles in Fig.~\ref{fig:questions-1-2-1}. On the other hand, an inspection on the evolution of the  projected dynamics onto each POD mode reveals that the G-ROM is performing actually quite well in modeling the dynamics of the first two modes, while its performance deteriorates for higher frequency modes; see Fig.~\ref{fig:GROM_projection} for the case $r=6$. 

For the SBE problem studied here, as the dimension of the G-ROM increases, the overall accuracy also improves, as can already be seen in Fig.~\ref{fig:questions-1-2-1}. 
Note also that the G-ROM performance improves as the diffusion coefficient $\nu$ increases (results not shown). This behavior is expected since increasing $\nu$ increases the diffusion effects, which, in turn, reduces the steepness of the localized internal layer. These numerical results suggest the convection-dominated regime to be a primary cause of the G-ROM's numerical instability observed here, just as in the deterministic case~\cite{wells2016regularized}.

\subsection{L-ROM Results: Alleviation of G-ROM's Spurious Oscillations}
	\label{sec:question-3}

In this section, we illustrate that the G-ROM's spurious oscillations such as those illustrated in the previous section can be alleviated by using the L-ROM~\eqref{eqn:l-rom-3} derived in Section~\ref{sec:l-rom} when the spatial filtering parameter $\delta$  is appropriately calibrated; cf.~\eqref{eqn:df_phi}.

\begin{figure}[h]
	\includegraphics[height=0.4\textwidth,width=1\textwidth]{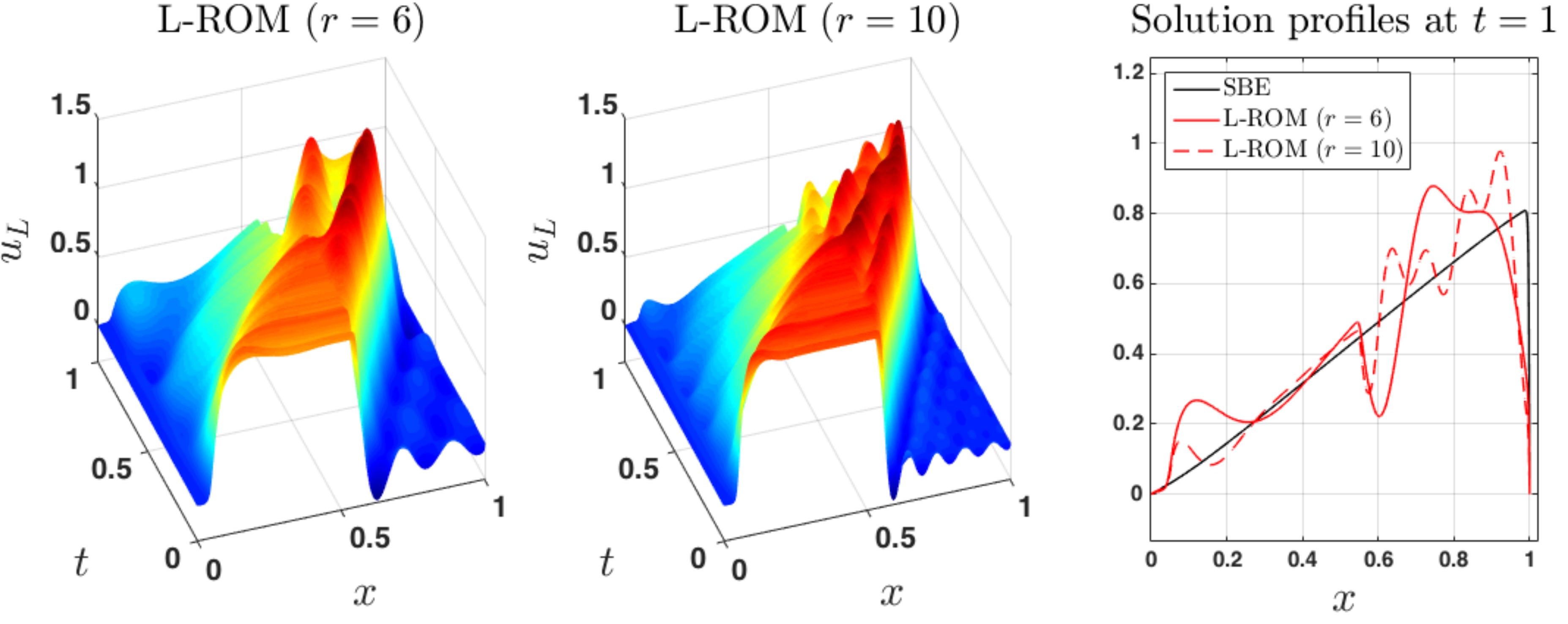}	
	\caption{
		Results corresponding to Fig.~\ref{fig:questions-1-2-1}  for the L-ROM~\eqref{eqn:l-rom-3}, where the spatio-temporal field $u_L:=\sum_{j=1}^r a_j \varphi_j$ is reconstructed from the numerical simulation of \eqref{eqn:l-rom-3} with dimension $r=6$ (left panel) and $r=10$ (middle panel).
	}
	\label{fig:question-3-1} 
\end{figure}

\begin{figure} 
\begin{center}
	\includegraphics[height=0.45\textwidth,width=1\textwidth]{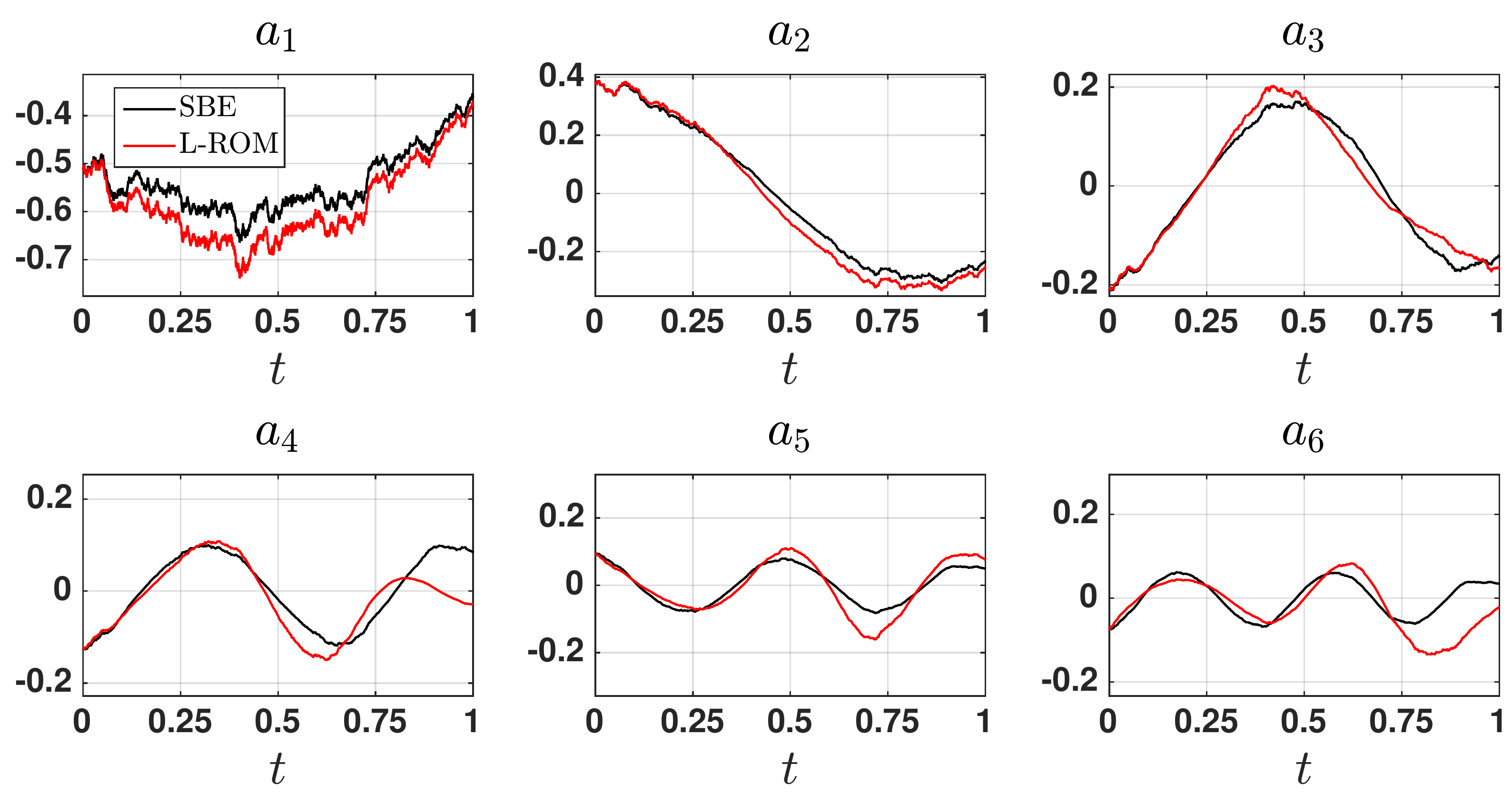}	
	\caption{
	The time series $a_j$, $1\le j \le r$, as modeled by the L-ROM~\eqref{eqn:l-rom-3} with dimension $r=6$ (red curves). Also plotted are the SBE solution projections onto the first $r$ POD modes (black curves).
	}
	\label{fig:LROM_projection} 
	\end{center}
\end{figure}

We choose the optimal value of this free parameter $\delta$ to be the value that minimizes the $L^2$-error of the corresponding L-ROM in reconstructing the SBE's spatio-temporal field. In our numerical experiments, we find the optimal value by trial and error.   To reduce the numerical efforts, especially in view of the statistical test given in the next section, all the numerical results related to the L-ROM~\eqref{eqn:l-rom-3} are obtained for $\delta = 0.12$, which is a nearly optimal $\delta$ value for the $r=10$ and $\sigma = 0$ case.\footnote{We have checked that, under the parameter setting used to generate Figs.~\ref{fig:question-3-1} and \ref{fig:LROM_projection}, the $\delta$ value we chose ($\delta = 0.12$) is close to the optimal $\delta$ values for both  the $r=6$ and $r=10$ cases.} The L-ROM results corresponding to those plotted in Figs.~\ref{fig:questions-1-2-1} and \ref{fig:GROM_projection} for the G-ROM are shown in Figs.~\ref{fig:question-3-1} and \ref{fig:LROM_projection}, respectively. As can be observed from these results, the spurious oscillations are indeed significantly reduced in the L-ROM dynamics, and an improvement in the modeling of the SBE's spatio-temporal field is also achieved. 

It is also interesting to note that although the regularization used in the L-ROM successfully reduces the spurious oscillation observed in the G-ROM's high-frequency modes, it leads to a slight deterioration on the modeling of the projected dynamics onto the first POD mode as can be seen by comparing the upper left panels of Fig.~\ref{fig:LROM_projection} and Fig.~\ref{fig:GROM_projection}. This deterioration is also observed even if the optimal $\delta$ value is used. Of course, the deterioration is reduced when the dimension of the L-ROM is increased. We intend to further investigate this issue (together with potential L-ROM improvements) in a separate communication. 

    \subsection{Robustness of the L-ROM results}  
	\label{sec:question-4}

In this section, we present some further numerical results regarding the statistical relevance of the results given in Sections \ref{sec:questions-1-2} and \ref{sec:question-3}. We also investigate the effect of the magnitude of the noise on the results. 

For this purpose, the performances of the G-ROM and L-ROM are assessed by using the relative $L^2$-errors computed as follows:
\begin{eqnarray} \label{eqn:rel_error}
	E(\omega)
	=  \frac{\sqrt{\int_{0}^1\int_{0}^1 |u(\cdot,\cdot; \omega) - u_r(\cdot,\cdot; \omega)|^2 \d x \d t}}{\sqrt{\int_{0}^1\int_{0}^1 |u(\cdot,\cdot; \omega)|^2 \d x \d t}} \times 100 \%,
\end{eqnarray}
where for each sample path $\omega$, $u(\cdot,\cdot; \omega)$ denotes the solution to the SBE \eqref{eqn:sbe}, and  $u_r(\cdot,\cdot; \omega)$ denotes the solution to either the G-ROM~\eqref{eqn:g-rom-2} or the L-ROM~\eqref{eqn:l-rom-3} with dimension $r$. 

\begin{figure}
\begin{center}
	\includegraphics[height=0.35\textwidth,width=0.8\textwidth]{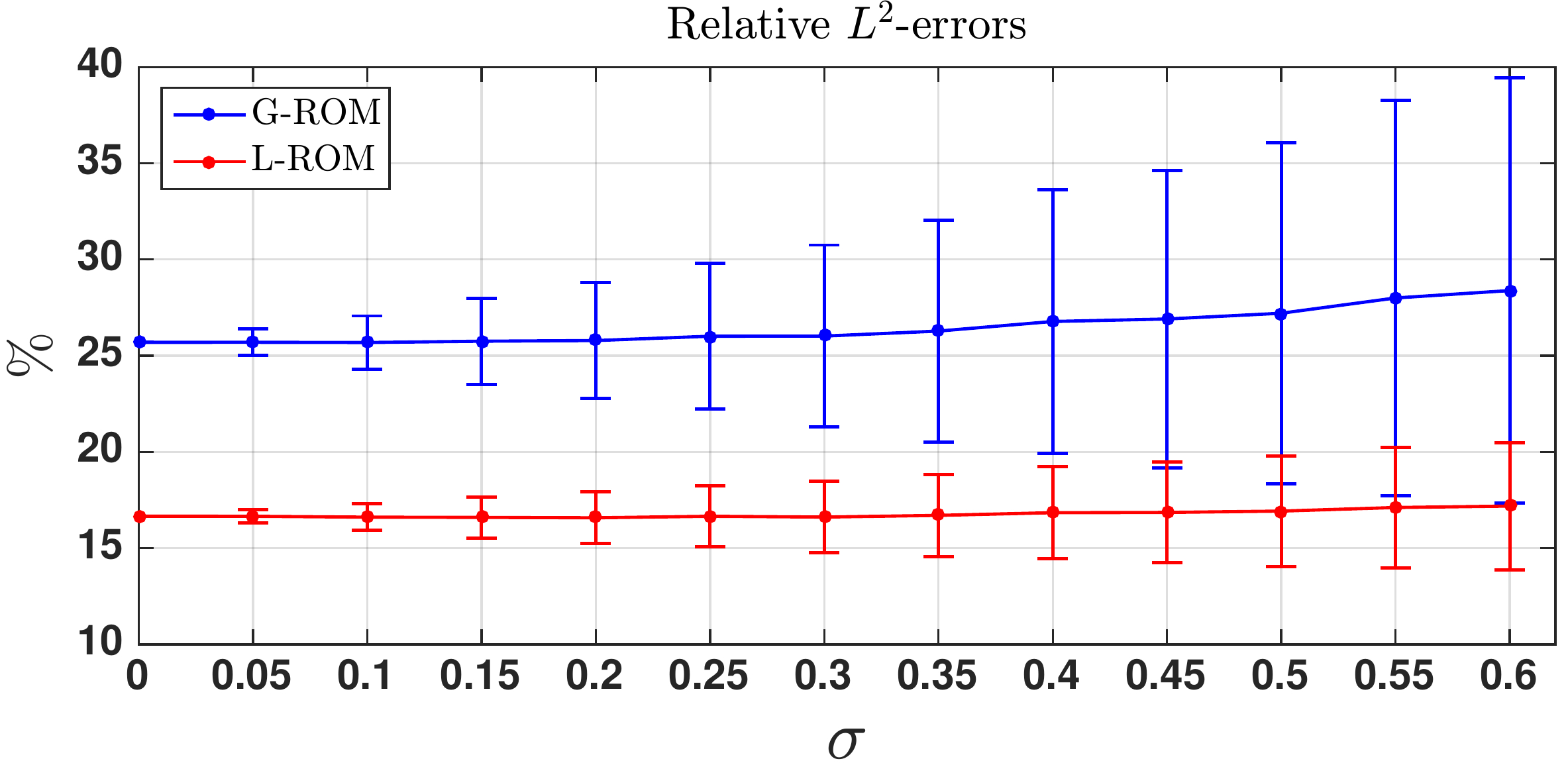} 
	\caption{
		Relative $L^2$-errors associated with the G-ROM~\eqref{eqn:g-rom-2} and the L-ROM~\eqref{eqn:l-rom-3} as computed via \eqref{eqn:rel_error} for $r=10$. The errors are computed for 13 values of the noise amplitude parameter $\sigma$ equally spaced between $0$ and $0.6$. An ensemble simulation of size $3000$ is carried out for the SBE~\eqref{eqn:sbe} and the two ROMs \eqref{eqn:g-rom-2} and \eqref{eqn:l-rom-3} for each $\sigma$ value. The ensemble averages of the relative errors are plotted. The error bars indicate the standard deviations. The parameter $\delta$ used in the differential filter involved in the L-ROM \eqref{eqn:l-rom-3} is fixed to be $0.12$ for all the simulations. 
	}
	\label{fig:question-4-1} 
	\end{center}
\end{figure}

We consider 13 noise magnitude $\sigma$ values equally spaced between $0$ and $0.6$.  For each of these $\sigma$ values, we perform $3000$ numerical simulations of the fine resolution discretization of the SBE (to obtain $u$) and the two ROMs (to obtain $u_r$). The dimension of the ROMs is chosen to be $r=10$, and the parameter $\delta$ used in the differential filter involved in the L-ROM \eqref{eqn:l-rom-3} is fixed to be $0.12$ (cf.~Section~\ref{sec:question-3}). In Fig.~\ref{fig:question-4-1}, the ensemble averages of the relative errors are plotted; the error bars indicate the standard deviations.\footnote{We checked that the statistical results plotted in Fig.~\ref{fig:question-4-1} have already converged by comparing the results estimated from 1500 sample points of the relative errors for each of the $\sigma$ values.} This figure shows that the L-ROM is not only more accurate but also more robust to noise variations than the G-ROM. Indeed, for the larger $\sigma$ values  in Fig.~\ref{fig:question-4-1}, the standard deviations of the relative $L^2$-errors associated with the G-ROM are significantly larger than those associated with the L-ROM as indicated by the corresponding error bars.

Finally, we mention that for the simulation of the G-ROM and L-ROM, instead of updating the POD basis for each $\sigma$ value and for each realization of the noise, we have fixed the POD basis to be the one constructed from the spatio-temporal field of the SBE at $\sigma=0$ (i.e., the deterministic Burgers equation). We made this choice based on the observation that the POD bases for different $\sigma$ (within the explored range $[0,0.6]$) and different noise paths actually resemble the POD basis for the $\sigma=0$ case, which is a feature that is specific to the linear multiplicative noise. When the POD basis is updated for each noise path and each $\sigma$,  we obtain results that are similar to those plotted in Fig.~\ref{fig:question-4-1}, although the standard deviations of the G-ROM errors are slightly reduced and the standard deviations of the L-ROM errors are slightly increased. 

\section{Conclusions and Outlook}
	\label{sec:conclusions}

Numerical instability is a significant challenge for standard ROMs of deterministic convection-dominated fluid flows.
A natural question is how (if at all) this challenge translates to ROMs of stochastic fluid flows.
In this paper, we took a modest step toward investigating this question by performing a computational study of the SBE~\eqref{eqn:sbe} with a small diffusion coefficient ($\nu = 10^{-3}$) and in the presence of a steep internal layer.
The numerical results suggested that standard (Galerkin) ROMs display spurious numerical oscillations in this convection-dominated regime.
To alleviate these oscillations, we tested the L-ROM, which is a regularized ROM that uses explicit spatial filtering to smooth (regularize) the convective term in the SBE.
The L-ROM results were significantly more accurate than the G-ROM results.
In particular, the G-ROM numerical oscillations were significantly decreased by the L-ROM; compare Figs.~\ref{fig:questions-1-2-1} and \ref{fig:GROM_projection} with Figs.~\ref{fig:question-3-1} and \ref{fig:LROM_projection}. Furthermore, the L-ROM results were less sensitive to noise magnitude variations than the G-ROM results; see Fig.~\ref{fig:question-4-1}.

We emphasize that much more remains to be done for a clear understanding of the potential numerical instability of ROMs and possible remedies of such instability for convection-dominated stochastic flows. 
For example, it is interesting to explore whether the results of this computational study are valid for other types of noise (e.g, additive noise or correlated additive and multiplicative noise) and more realistic settings (e.g., 3D fluid flows modeled by the NSE). Furthermore, it is also interesting to investigate the performance of other regularized ROMs (e.g., the evolve-then-filter ROM~\cite{wells2016regularized}) and stabilized ROMs. 

Another important research direction is the investigation of the robustness of the proposed ROMs.
For example, it would be interesting to use more systematic approaches (such as numerical analysis~\cite{giere2015supg}) to determine general scalings for the model parameters, such as the spatial filtering parameter $\delta$ used in the L-ROM.
One could also perform thorough sensitivity studies of these ROMs with respect to model parameters, such as the number of basis functions ($r$) or the filtering parameter ($\delta$).

\section*{acknowledgement}

The authors greatly appreciate the financial support of the National Science Foundation through grants DMS-1522656 (TI and XX) and DMS-1616450 (HL). We would also like to thank Prof. Gianluigi Rozza for bringing to our attention reference \cite{Torlo16}.


\bibliographystyle{plain}
\bibliography{traian,Stoch_ROM}
\end{document}